# Thermal modification of ZrCu metallic glass nanolaminates: Structure and mechanical properties

Andrea Brognara[1], Chanwon Jung[1,a)], Cristiano Poltronieri[2,b)], Philippe Djemia[2], Gerhard Dehm[1,*], Matteo Ghidelli[2,*] and James P. Best[1,*]

[1] *Max Planck Institute for Sustainable Materials, Max-Planck-Str. 1, 40237 Düsseldorf, Germany*

[2] *Laboratoire des Sciences des Procédés et des Matériaux (LSPM), CNRS, Université Sorbonne Paris Nord, 93430 Villetaneuse, France*

[a)] *Present address: Department of Materials Science and Engineering, Pukyong National University, 45 Yongso-ro, Nam-gu, 48513 Busan, Republic of Korea*

[b)] *Present address: Carl Zeiss S.p.A., Via Varesina 162, 20156 Milan, Italy*

[*] *Corresponding authors*: j.best@mpie.de (J.P. Best), matteo.ghidelli@lspm.cnrs.fr (M. Ghidelli), dehm@mpie.de (G. Dehm)

**Abstract**

The effects of thermal treatments on metallic glass nanolaminates (NLs), with a composition of $Zr_{24}Cu_{76}$ and $Zr_{61}Cu_{39}$ and a bilayer period of 50 nm, were explored to control their mechanical properties through annealing-induced atomic structure modifications, structural relaxation, and partial crystallisation. Annealing treatments up to 330 °C ($T < T_g$, the glass transition temperature) maintain the amorphous structure of the NLs, while inducing atomic structural relaxation, densification, and free volume annihilation, reducing the formation of corrugations on fracture surfaces. Atom probe tomography measurements reveal that annealing at 330 °C for 60 mins also causes intermixing between layers, altering their compositions to $Zr_{44}Cu_{56}$ and $Zr_{55}Cu_{45}$ with increased mixing enthalpy. Moreover, the NLs exhibit superior thermal stability against crystallisation compared to their monolithic counterparts, remaining amorphous up to 420 °C, while the monolithic $Zr_{24}Cu_{76}$ and $Zr_{61}Cu_{39}$ films are crystalline at 390 °C, as a result of chemical interdiffusion and the heterogeneous NL structure of delaying the onset of crystallisation. Annealing treatments $T > T_g$ (~420 °C) induce partial crystallisation, forming Cu-Zr-based intermetallic and Zr-oxide phases, whereas annealing $T < T_g$ (330 °C) retains a visible layer structure. Nanoindentation analyses show a progressive increase in elastic modulus and hardness for higher annealing temperatures as a result of structural relaxation and likely nanocrystal formation, with a maximum hardness equal to 7.6 ± 0.2 GPa obtained after heat treatment at 420 °C for 60 mins and exceeding the rule-of-mixtures. These results highlight the potential of thermal treatments to tailor the structural, mechanical and thermal properties of metallic glass nanolaminates for advanced material applications.

## 1. Introduction

An active key research topic for bulk and thin film metallic glasses (BMGs and TFMGs, respectively) is the control of local heterogeneities aimed to mitigate macroscopically brittle behaviour via shear band (SB) localisation [1-3]. One common strategy is to perform annealing treatments, thereby modifying the mechanical properties and deformation behaviour [4, 5]. The effects vary depending on whether the annealing is performed below or above the glass transition temperature ($T_g$). Annealing below $T_g$ induces local atomic rearrangements known as structural relaxation, leading to densification and annihilation of free volume [6, 7]. These processes enhance mechanical properties such as hardness ($H$) and fracture strength [6, 8-10]. For example, annealing a ZrCu-based BMG below $T_g$ increased $H$ from ~6.5 up to 7.3 GPa and elastic modulus ($E$) from around 113 up to 131 GPa [6], while similar results are obtained for $Zr_{40}Cu_{60}$ TFMG showing an increment of $H$ and $E$ from 6.1 up to 6.5 GPa and from 100 up to 109 GPa, respectively [8]. However, this reduction of free volume also decreases the plastic deformability of metallic glasses [11, 12] due to reduced nucleation sites for shear bands (SBs), resulting in lower fracture strain [5, 12, 13].

In contrast, annealing at $T > T_g$ can cause the nucleation of nanocrystals within the amorphous matrix. These nanocrystals can act as barriers against SB propagation, enhancing hardness and potentially increasing plasticity in ZrCu-based alloys [14, 15]. For instance, annealing ZrCu-based BMGs above $T_g$ increased fracture limit during compression from ~2% to ~10% plastic strain [14]. Molecular dynamics (MD) simulations have also shown that the nucleation of nanocrystals softer than the glass matrix can induce a toughening effect [16], which is more pronounced for relatively small crystallites <2 nm. Nevertheless, literature also reports embrittlement phenomena as result of the precipitation of harder nanocrystals with intermetallic compositions, which can induce embrittlement due to the development of detrimental stress fields in the material [5, 13].

Another common approach to control local heterogeneities is represented by the synthesis of nanostructured metallic glasses [17, 18], and fully amorphous or amorphous/crystalline nanolaminates (NLs) [19-21], in improving mechanical behaviour. Specifically, the presence of regions with different local chemistry and atomic order, the formation of nanoclusters as well as the high density of interfaces between nanolayers with large chemical/structural contrast [22-24], has been shown to inhibit the formation of catastrophic single SB events, while enhancing mechanical properties [25-28]. In this context, annealing treatments can be exploited to further tune the degree of heterogeneity by introducing nanocrystals and promoting interface diffusion, representing an effective strategy to tailor and boost the mechanical properties, while mitigating SB localisation. As an example, $Sc_{75}Fe_{25}$ nanoglasses exhibited up to ~10% plastic deformability during micro-compression and increased yield stress from ~1.2 up to ~1.6 GPa after annealing at 200 °C, due to the presence of high free volume at glass-glass interfaces [9]. Moreover, cluster-assembled ZrCu/O TFMGs reported lower thermal stability than their compact counterparts, while annealing treatment at $T > T_g$ led to an increment in $E$ and $H$ up to 180 and 14 GPa, respectively, together with the development of residual stresses reaching 1480 MPa at 420 °C [29]. However, the effects of thermal treatments on the structure and mechanical properties of heterogeneous glassy materials are not fully understood and require further investigation.

In this study, we focus on the effects of annealing on $Zr_{24}Cu_{76}/Zr_{61}Cu_{39}$ amorphous NLs, with bilayer period ($\Lambda$) equal to 50 nm deposited by magnetron sputtering. Such bilayer period has been selected because the larger plasticity (up to 16% strain in compression [21]) originated by the high density of nanointerfaces compared to the brittle NL with $\Lambda = 200$ nm. Additionally, the relatively larger layer thickness compared to NL with $\Lambda = 25$ nm made it easier to evaluate morphological, structural, and chemical evolutions. Here, we evaluate the resistance against crystallisation of the NL compared to its monolithic constituents, focusing on understanding annealing-induced structural changes, and the promotion of partial crystallisation to impede SB propagation and improve mechanical properties. The atomic- and micro-structural evolution are explored through X-ray diffraction (XRD), electron microscopy, and atom probe tomography (APT) analyses, unravelling structural evolution with annealing temperature. In parallel, we carried out nanoindentation measurements, investigating the evolution of mechanical properties.

## 2. Materials and methods

### 2.1. Synthesis of fully-amorphous nanolaminates

ZrCu amorphous nanolaminates (NLs) were synthesised by magnetron sputtering, as detailed in Ref. [21]. The composition of the individual layers (*i.e.* $Zr_{24}Cu_{76}$ and $Zr_{61}Cu_{39}$, at.%) was controlled by adjusting the power ratios applied to the pure Cu and Zr targets [22]. An automated deposition process was employed, varying the sputtering power and alternately opening and closing both the substrate and cathode shutters to control the thickness and composition of each layer. The NLs have a total thickness of 3 μm and bilayer period ($\Lambda$) equal to 50 nm. The $Zr_{24}Cu_{76}$ layer was always in contact with the substrate, whereas the $Zr_{61}Cu_{39}$ layer was always positioned at the top surface of the NLs.

### 2.2. Annealing treatments

To investigate the effects of annealing on the amorphous NL structure and mechanical properties, annealing treatments at various targeted temperatures were conducted. The primary objectives were: (*i*) Anneal the NL at sufficiently low temperature to avoid crystallisation ($T < T_g$) to assess the presence of intermixing between the layers; (*ii*) Induce partial crystallisation ($T > T_g$) to determine if the crystallites could act as additional barriers against SB propagation [6, 14, 15]. Annealing treatments were performed at various temperatures ranging from 330 to 450 °C. The annealing was conducted in a vacuum chamber at a pressure of $5 \times 10^{-3}$ Pa. Both the heating and cooling ramps were set at 10 °C/min. The annealing time was 60 minutes for all temperatures, with the addition of 80 and 120 minutes for annealing at 420 °C.

*2.3. Structural characterisation*

The thickness and morphology of the NLs were inspected using scanning electron microscopy (SEM, Zeiss Auriga dual-beam). To examine the morphology evolution of the layer stack caused by annealing, a dual-beam SEM using a gallium focused ion beam was used to prepare cross-sectional cuts of the specimens. Structural analysis of the films was conducted using grazing incidence X-ray diffraction (GIXRD) with Cu Kα radiation ($\lambda = 0.154$ nm), utilising a Rigaku SmartLab 9 kW diffractometer. Measurements were carried out at 45 kV and 200 mA, with XRD data collected over a 2θ range of 20°–100° in continuous scan mode, with a step size of 0.01°, and a scanning speed of 0.75°/min.

Thermal stability and the onset of crystallisation were assessed by acquiring *in situ* XRD spectra while increasing the temperature from $T_{room}$ to 600 °C [22]. To prevent thermally activated interdiffusion and Zr-Si reactions, the films were deposited on Si (100) wafers coated with an amorphous $SiN_x$ diffusion barrier layer. The heating rate was set to 30 °C/min, and diffractograms were recorded at constant temperatures every 30 °C with holding times of ~20 min. The *in situ* XRD heating was conducted in a He atmosphere ($1.35 \times 10^5$ Pa) to limit oxidation. These measurements were performed using the same setup as for the as-deposited films but in Bragg-Brentano geometry (θ/2θ scan axis) with a reduced range of 25°-90°, scanned in continuous mode with a step size of 0.01° and a scanning speed of 3°/min. An offset of 3° was applied to avoid interference from the Si substrate signal.

APT was used to investigate the chemical profile of NLs and the fluctuations within shear bands, using volumes of material deformed by nanoindentation. The APT specimens were prepared using a Thermo Fisher Scios 2 dual-beam FIB-SEM, following the fabrication protocol in Ref. [30]. APT analyses were performed using a local electrode atom probe (CAMECA LEAP 5000 XS) in pulsed laser mode at a specimen base temperature of 50 K. A laser pulse energy of 40 pJ, a detection rate of 1%, and a laser pulse frequency of 125 kHz were used. Data reconstruction and analyses were carried out with the AP Suite 6.3 software (CAMECA Instruments).

*2.4. Mechanical analyses*

Nanoindentation measurements were conducted using a continuous stiffness measurement (CSM) configuration under load control with a fixed load rate set to 0.05 s$^{-1}$. A KLA G200

Nanoindenter DCM head equipped with a diamond Berkovich tip (Synton-MDP, Switzerland) was used, following an experimental procedure similar to that described in Ref. [22]. Hardness ($H$) and elastic modulus ($E$) values were extracted using the Oliver and Pharr method [31], for depths below approximately 10% of the film thickness to avoid the influence of the Si (100) substrate. Brillouin light scattering (BLS) experiments have been carried out using an identical methodology to that reported in Ref. [21] and considering NLs anisotropy. For both the as-deposited and sample annealed at 330 °C, the elastic modulus was determined from the shear modulus using a Poisson's ratio of 0.38, while no signal was obtained for samples annealed over 400 °C.

## 3. Results and discussion

### 3.1. Effect of nanolayering on thermal stability and structural evolution

The evolution of the NL amorphous structure as a function of temperature was characterised through *in situ* XRD heating as already performed on monolithic ZrCu TFMGs in Ref. [22]. Fig. 1a shows the XRD diffractograms of the NL at increasing temperatures. At $T_{room}$, the amorphous peak of the NL corresponds to the superposition of the peaks of the individual $Zr_{24}Cu_{76}$ and $Zr_{61}Cu_{39}$ layers (with maxima at 41.0° and 36.6° respectively), which constitute the NL stack (see Fig. S1 in the Supplementary Information). The amorphous peak remains stable up to 270 °C; however, at 300 °C, the two superimposed peaks merge into a single amorphous peak that remains visible up to 420 °C. From 450 °C onwards, the NL shows a fully crystalline structure consisting of several Cu-Zr intermetallics and Zr-oxide phases (see Fig. 1a).

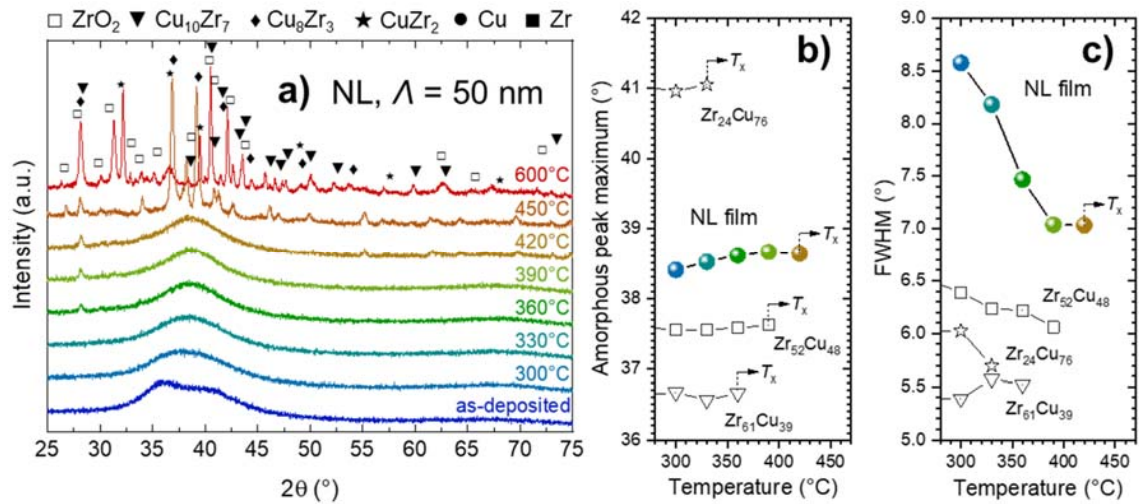

**Fig. 1**: Structural evolution during *in situ* XRD heating. (a) At $T_{room}$ the NL exhibits an amorphous peak resulting from the superposition of its two constituent layers, $Zr_{24}Cu_{76}$ and $Zr_{61}Cu_{39}$. Starting form 300

°C, the two peaks merge due to chemical intermixing. The first crystallisation peaks appear at 360 °C, but the NL maintains its amorphous structure up to 420 °C, with full crystallisation occurring at 450 °C. The evolution of the amorphous peak position (b) and FWHM (c) of the NL after the peaks merge (300 °C) indicates structural and chemical changes in the NL before crystallisation at temperature $T_x$. The NL exhibits greater resistance to crystallisation compared to the monolithic compositions [22].

The merging of the two superimposed amorphous peaks at 300 °C is due to an evolution of the local atomic structure, suggesting chemical intermixing between the layers before crystallisation occurs [32]. The structural evolution of the NL is confirmed by the shift towards higher 2θ of the position of the maximum of the merged amorphous peak between 300 °C and 420 °C (*i.e.* from 38.4 up to 38.7°, Fig. 1b) and the decrement of the full width at half maximum (FWHM) from ~8.6 to ~7° (Fig. 1c). This suggests a decrease in the average atomic distances (*d*) of ~0.02 Å, calculated using the Ehrenfest equation ($2d \cdot sin(\theta) = 1.23\lambda$) as in Refs. [22, 33]. The position of the maximum of the NL merged peak at 300 °C (*i.e.* 38.4°) falls halfway between those of the $Zr_{24}Cu_{76}$ and $Zr_{61}Cu_{39}$ monolayers (*i.e.* 41.0° and 36.6°, respectively) and corresponds to the peak position of an ideal monolithic film with an average composition between those of the layers: approximately $Zr_{50}Cu_{50}$ [22, 34].

Remarkably, the NL exhibits higher resistance to crystallisation compared to its monolithic counterparts (*i.e.* $Zr_{24}Cu_{76}$ and $Zr_{61}Cu_{39}$ TFMGs), reaching a crystallisation temperature ($T_x$) of at least 390 °C (see Fig. 1a). Although the first crystalline peaks likely corresponding to $Cu_{10}Zr_7$ and $ZrO_2$ phases start to appear at 360 °C, the NL structure remains amorphous up to 420 °C, while the $Zr_{24}Cu_{76}$ and $Zr_{61}Cu_{39}$ monolithic films show a completely crystallised structure already at $T_x$ of 390 °C, as reported in Ref. [22].

The increased thermal stability of the NL could be related to atomic intermixing phenomena, which delay the crystallisation process, as reported in Ref. [32]. Initially, the heterogeneous structure with numerous interfaces activates diffusion processes to reduce the composition gradient between the $Zr_{24}Cu_{76}$ and $Zr_{61}Cu_{39}$ layers, evolving towards an average $Zr_{50}Cu_{50}$ composition. Secondly, $Zr_{50}Cu_{50}$ is known to have a larger thermal stability against crystallisation [22, 35, 36], further delaying the onset of crystallisation. The combination of these factors justifies the greater resistance to crystallisation of the NL compared to its individual components.

*3.2. Effect of isothermal annealing on structure and morphology*

Based on *in situ* XRD heating results from Fig. 1a, isothermal annealing experiments were performed within a temperature range starting at 330 °C, which induces structural evolution while avoiding crystallisation, and gradually increasing up to 450 °C to induce crystallisation and evaluate the process kinetics. The XRD diffractograms of the annealed NL at different temperatures are shown in Fig. 2. Due to differences in thermal profiles between XRD and isothermal annealing experiments, direct comparison of the results is not straightforward.

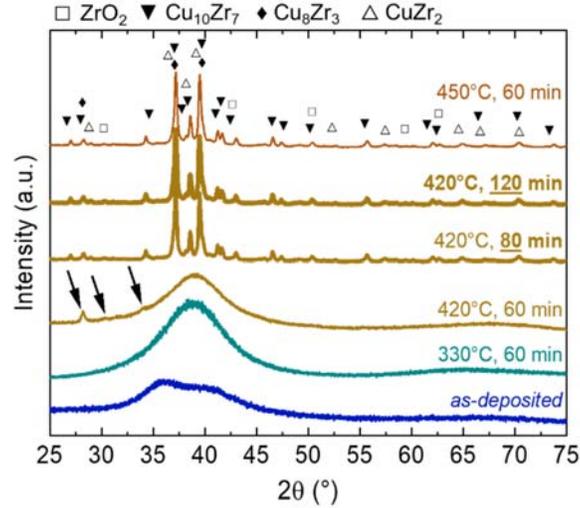

**Fig. 2**: Evolution of XRD diffractograms of the NLs at different annealing temperatures and durations. The sample annealed at 330 °C remains in an amorphous phase without crystallisation. However, the amorphous peak evolves from the superimposed shape of the as-deposited condition into a single peak. In contrast, annealing at 420 °C for 60 minutes shows the appearance of the first crystalline peaks mostly of oxides at lower $2\theta$ (highlighted by the arrows) while still retaining an amorphous matrix. Samples annealed for longer times and at $T \geq 420$ °C are completely crystalline.

Annealing at 330 °C for 60 minutes successfully induced structural evolution without crystallisation (Fig. 2). The XRD diffractogram maintained an amorphous structure with no evident crystalline phases, though the merging of the two amorphous peaks suggests structural changes. This indicates that 330 °C is sufficiently below the $T_g$ of the NL, aligning with *in situ* heating XRD analysis reported in Fig. 1. To achieve a composite nanocrystalline-amorphous structure and investigate the NL crystallisation process and kinetics, annealing treatments were conducted at $T > 330$ °C to exceed $T_g$.

At 420 °C for 60 minutes, the film structure remained primarily amorphous, but several peaks appeared (Fig. 2), corresponding to crystalline $Cu_{10}Zr_7$, $Cu_8Zr_3$ and $ZrO_2$ phases. This suggests that 420 °C exceeds the crystallisation temperature ($T_x$) of the NL's components, which were reported to show signs of crystallisation at 330 °C and 360 °C [22]. However, structural rearrangements at the interfaces may slow down crystallisation [32], allowing the NL to retain some amorphous phase at this temperature.

Annealing at 420 °C for 80 and 120 minutes resulted in full crystallisation of the films (Fig. 2); the XRD diffractograms show the presence of intermetallic $Cu_{10}Zr_7$, $Cu_8Zr_3$ phases together with ZrO$_2$, in agreement with Fig. 1. Further annealing at 450 °C produced a diffractogram identical to those of samples annealed at 420 °C for 80 and 120 minutes, indicating that full crystallisation already occurred.

The morphology and fracture surface of the NLs under different annealing conditions were also examined. Fig. 3 and 4 present the SEM and FIB cross-section images of samples with amorphous and completely crystallised structures. The as-deposited sample displays typical corrugations of MG fracture surfaces [11] with an average width of ~150 nm (Fig. 3a). After annealing at 330 °C and 420 °C for 60 minutes (Fig. 3c,e), the NLs still exhibit corrugations, but these are more uniformly distributed and have a smaller average width of ~60 nm, indicating extremely localised plasticity associated with (brittle) SB propagation [11, 37]. The average corrugation width is reported to correlate with the glass toughness [38-40]. A reduction in corrugation width is associated with a decrease in free volume content and a reduction of fracture toughness and ductility, as a result of annealing-induced structural relaxation [5, 12, 13].

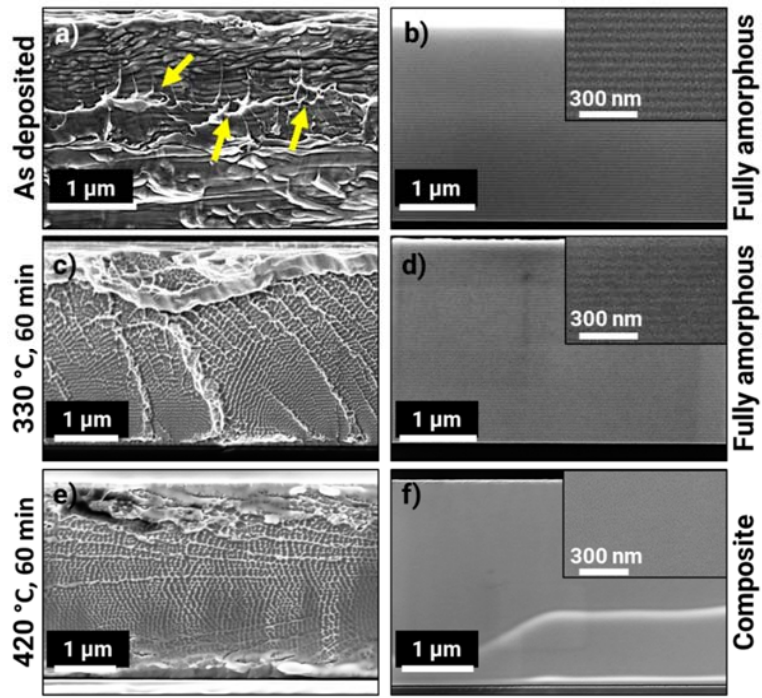

**Fig. 3**: SEM images of the fracture surfaces (*left*) and FIB cross-sections (*right*) of the as-deposited NL (a,b), and after annealing at (c,d) 330 °C for 60 min, and (e,f) 420 °C for 60 min. The as-deposited NL shows the presence of nanolayers (b) and corrugations, highlighted by the yellow arrows in (a). After annealing, the fracture surfaces do not show nanolayers (c,e), and the corrugations have a smaller width and spacing. For the sample annealed at 330 °C, the FIB cross-section (d) reveals that the layer structure is still intact, whereas for the sample annealed at 420 °C (f), it is not possible to distinguish any layers.

Despite the changes in fracture morphologies for samples annealed at 330 °C and 420 °C for 60 minutes, the presence of nanolayers becomes difficult to distinguish (Fig. 3c,e). FIB cross-section cuts revealed that the layer structure remains intact after annealing at 330 °C (Fig. 3d), but not after annealing at 420 °C for 60 minutes. Fig. 4 shows that the fracture surfaces of samples annealed at 420 °C for 80 minutes and 120 minutes and at 450 °C for 60 minutes are largely homogeneous, with FIB cross-sections revealing the presence of crystalline domains, and no evidence of nanolayers, consistent with their fully crystalline XRD diffractograms (Fig. 2).

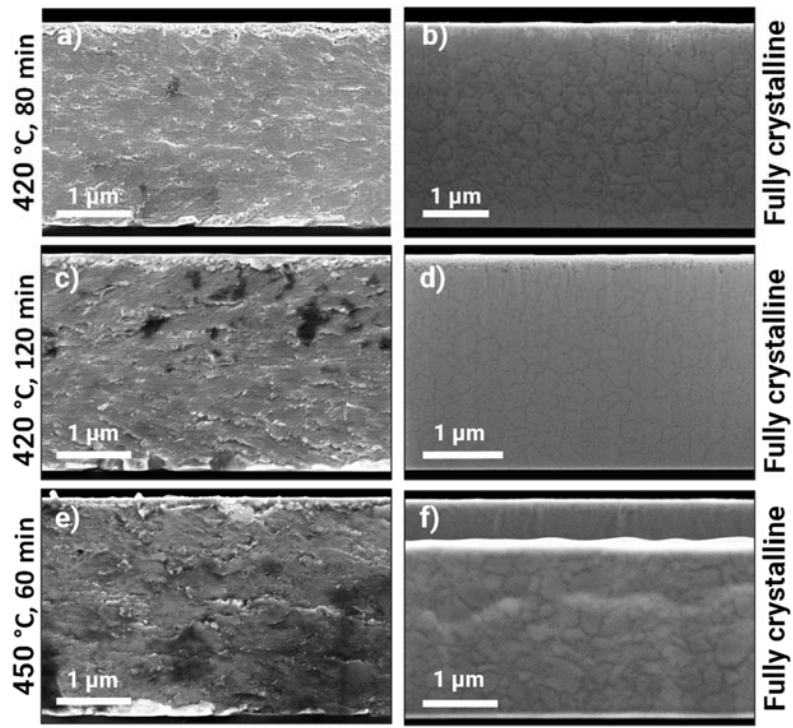

**Fig. 4**: SEM images of the fracture surfaces (*left*) and FIB cross-sections (*right*) of the NLs with a completely crystalline structure after annealing at (a,b) 420 °C for 80 min, (c,d) 420 °C for 120 min, and (e,f) 450 °C for 60 min.

APT analysis was conducted to investigate the evolution of the NL structure after annealing and verify chemical intermixing between the layers after annealing at 330 °C, complementing the *in situ* XRD analysis. Fig. 5 compares the APT analyses of the NL in the as-deposited condition and after annealing at 330 °C for 60 minutes. The 1D concentration profile of the annealed NL (Fig. 5d) shows a significant change in the composition of the layers compared to the as-deposited sample (Fig. 5c and Fig. S2 of the Supplementary Information). The content of Cu changed from ~70 and ~30 at.% for the layers of the as-deposited film, down to ~56 and 45 at.% after annealing (with balance Zr). Despite chemical intermixing and structural changes, separate layers remain (Fig. 5b), with largely unaltered individual thicknesses (Fig. 5a). Oxygen content analysis indicated no significant oxidation during annealing (Fig. 5e,f). The concentration of oxygen for both samples is higher in the Zr-rich layers due to the higher oxygen affinity of Zr and reaches a maximum value of ~0.5 at.%.

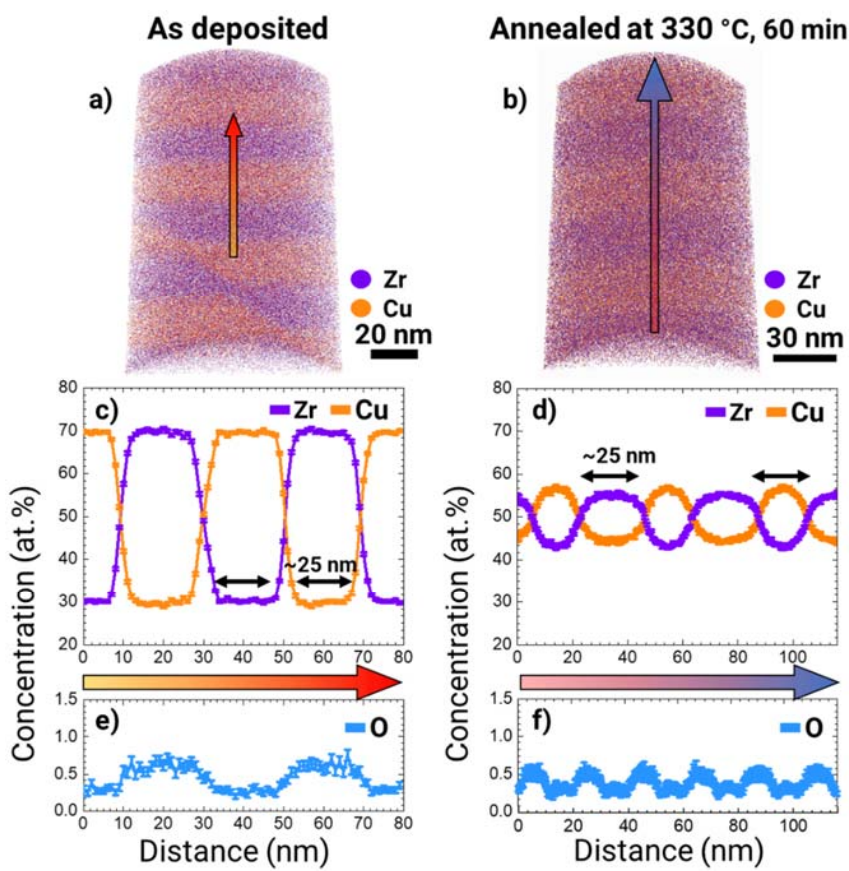

**Fig. 5**: Comparison of APT results for (a,c,e) as-deposited NL and (b,d,f) annealed at 330 °C for 60 min. 3D atomic maps of (a) as-deposited and (b) annealed NLs showing that layers remain visible even after thermal treatment. A shear band is captured in the volume of (a). The 1D concentration profile scans (indicated by colour gradient arrows) reveal differences in the compositions of the NL layers between the (c) as-deposited and (d) annealed samples, indicating that interdiffusion has occurred between the layers. Oxygen concentration profiles show a relatively low average content of ~0.5 at.%, both (e) before and (f) after annealing.

The evolution of the amorphous NL peak position and FWHM during *in situ* XRD heating suggests that thermally activated interdiffusion processes continue with increasing annealing temperature or duration (Fig. 1), aligning with APT data. The average atomic distance (*d*) for the XRD analysis is comparable to a monolithic TFMG with ~ $Zr_{50}Cu_{50}$ composition [22]. A $Zr_{50}Cu_{50}$ composition corresponds to the average of the layers before annealing, and generally has a higher thermal stability due to a lower relative value of the formation enthalpy $\Delta H$ [35, 36]. Therefore, the evolution of the layer chemistry towards compositions with higher crystallisation resistance would increase thermal stability of the amorphous phase NLs, as observed in Fig. 1. This supports the hypothesis that diffusion increases the thermal stability of the amorphous phase of the NL compared to its monolithic layer references. Further annealing at 420 °C for 60 minutes likely caused additional intermixing together with partial crystallisation, as observed in XRD and by the lack of visible layers in SEM micrographs.

*3.3. Effect of annealing temperature on mechanical properties*

Fig. 6 shows the evolution of the elastic modulus ($E$) and hardness ($H$) for the NL extracted by nanoindentation as a function of annealing temperature and duration. Fig. S3 of the Supplementary Information presents the evolution of properties against nanoindentation depth. Both $E$ and $H$ values increased with the annealing temperature, while the NL remained fully and partially amorphous (i.e., annealing at 330 and 420 °C for 60 min, respectively). From BLS measurements of the elastic modulus, similar trends were observed between the as-deposited and sample annealed at 330 °C. However, the $H$ value for the partially-amorphous sample annealed at 420 °C for 60 mins reached a higher value than that of the individual $Zr_{24}Cu_{76}$ and $Zr_{61}Cu_{39}$ sublayer constituents. Fully crystallised specimens (after annealing at 420 °C for 80 minutes) showed a further dramatic increment of $E$ and $H$ which then remained constant upon increases in annealing time or temperature. This is likely due to the increased $E$ and $H$ of the crystalline oxide and intermetallic phases formed in the crystallised sample. This confirms that full crystallisation was achieved, as also suggested by the lack of significant differences in XRD diffractograms (Fig. 2) and cross-section morphologies (Fig. 4) for samples annealed at 420 °C for 80 and 120 minutes and 450 °C for 60 minutes.

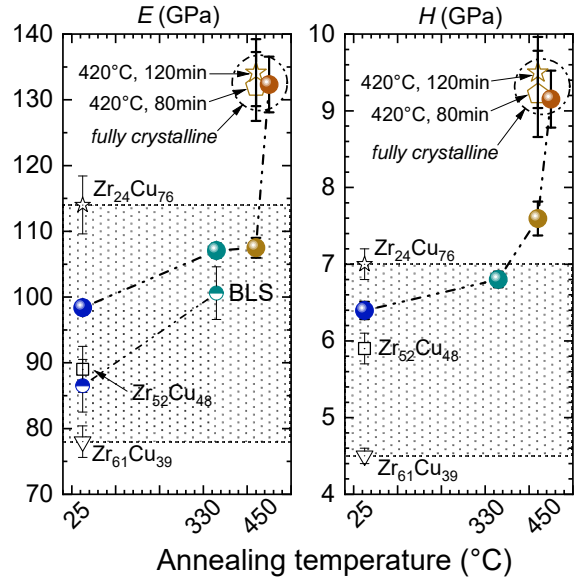

**Fig. 6**: Evolution of elastic modulus (*E*) and hardness (*H*) as a function of annealing temperature and time for the 3 μm thick NLs. The dotted region signifies that range between the *E* and *H* values of the as-deposited $Zr_{24}Cu_{76}$ and $Zr_{61}Cu_{39}$ monolithic references, respectively, from Ref. [22]. BLS values are also highlighted using half-filled circular symbols.

The increase in mechanical properties of the films that remained fully or partially amorphous is related to local chemistry and atomic rearrangements. As previously mentioned, the mechanical properties of TFMGs are closely related to their composition [22-24, 41]. However, in this case, the increase in *E* and *H* values cannot be attributed solely to changes in layer composition. This is because during annealing, the films tend towards an average $Zr_{50}Cu_{50}$ composition, which would show *E* and *H* values between those of the $Zr_{24}Cu_{76}$ and $Zr_{61}Cu_{39}$ monolithic references determined in Ref. [22], not higher as reported in Fig. 6, which suggest structural differences occurring from thermal treatment.

Therefore, the increase in mechanical properties must be related to other factors during annealing, including structural relaxation or the nucleation of crystallites [6, 9, 10]. The hypothesis of the annihilation of free volume during annealing at 330 °C is supported by the presence of corrugations with a smaller width on the NL's fracture surface compared to the as-deposited NL (see Fig. 3). The annihilation of free volume in metallic glasses is linked to a transition towards a more densely packed atomic configuration, resulting in higher stiffness [7, 22, 41]. Additionally, free volume reduction decreases the available sites for SB nucleation, increasing resistance to plastic flow and consequently inducing higher hardness [5, 6, 9, 10]. Furthermore, the presence of nanocrystals within the amorphous matrix in metallic glasses can increase hardness by acting as obstacles to SB propagation and increasing resistance to deformation [6, 42, 43].

Finally, these results can be compared with those of Apreutesei *et al.*, who performed annealing on monolithic $Zr_{60}Cu_{40}$ TFMGs at both $T < T_g$ and $T > T_g$ [8]. Similarly, annealing at $T < T_g$ induced structural relaxation without causing crystallisation, while partial crystallisation occurred when $T > T_g$, as reported for the ZrCu NL (Fig. 2). The thermal treatments and structural modifications led to increased $E$ and $H$ values for both the Cu-rich monolithic film and the NLs, as shown in Table 1.

**Table 1**: Comparison of nanoindentation properties for the $Zr_{24}Cu_{76}/Zr_{61}Cu_{39}$ NL of the present study and that of a monolithic $Zr_{60}Cu_{40}$ TFMG from Ref. [8], reporting their evolution for the as-deposited condition, fully amorphous (annealing $T < T_g$) and partially amorphous/crystallised (annealing $T > T_g$). $E$ values from BLS are included as $E_{BLS}$ for comparison. * value taken from Ref. [44] on the same material.

|  | Nanolaminate (NL), *this work* | | | Monolithic $Zr_{60}Cu_{40}$ from Ref. [8] | | |
| --- | --- | --- | --- | --- | --- | --- |
|  | *as-deposited* | $T < T_g$ (330 °C) | $T > T_g$ (420 °C) | *as-deposited* | $T < T_g$ (280 °C) | $T > T_g$ (380 °C) |
| $E$ (GPa) | 98 ± 1 | 107 ± 1 | 108 ± 2 | 100 ± 1 | 109 ± 1 | 115 ± 2 |
| $E_{BLS}$ (GPa) | 87 ± 4 | 101 ± 4 | - | 70 ± 1 * | - | - |
| $H$ (GPa) | 6.4 ± 0.1 | 6.8 ± 0.1 | 7.6 ± 0.2 | 6.1 ± 0.1 | 6.5 ± 0.1 | 6.6 ± 0.1 |

When the NL and monolithic samples had comparable structural conditions – namely as-deposited (*i.e.*, fully amorphous), fully amorphous after annealing at $T < T_g$, and partially crystalline after annealing at $T > T_g$ – they exhibited very similar $E$ values while the NL always had slightly higher $H$ values (*i.e.*, 6.6 GPa and 7.6 GPa for the monolithic film and the NL, respectively). These results can be put into context, considering that the mechanical properties of ZrCu TFMGs increase with their Cu content (at.%) [22, 41, 45], given that after annealing the NL tends towards an average composition of ~ $Zr_{50}Cu_{50}$ (see Fig. 5). This demonstrates that by exploiting the heterogeneous nanostructure of amorphous NLs, it is possible to obtain films with greater hardness, elastic modulus, and thermal resistance to crystallisation compared to monolithic TFMGs, highlighting that simple control over amorphous nanolayer architecture can play a significant role on the thermal stability and mechanical properties of thin-films.

## 4. Conclusions

This study focusses on the effects of thermal treatments on metallic glass nanolaminates (NLs) with a bilayer period ($\Lambda$) equal to 50 nm. The objective was to utilise annealing treatments to control the mechanical properties of the NLs by tuning the atomic structure, inducing structural relaxation, or partial crystallisation. The following conclusions can be drawn:

*Structural evolution and mechanical properties*

- Annealing induced a gradual evolution in the NL's amorphous structure and chemistry, with progressive increase in nanoindentation properties with increasing annealing temperatures. Up to 330 °C, the structure remained X-ray amorphous, but underwent structural relaxation, as indicated by the evolution of the amorphous peak shape (FWHM, and maximum position), suggesting densification, free volume annihilation, and changes in local order. The width of corrugations on the fracture surface were also reduced by annealing below $T_g$.
- Annealing caused intermixing between different layers and a variation in their individual compositions. The atomic composition in the NLs tended towards $Zr_{50}Cu_{50}$ after annealing at 330 °C.
- Annealing above $T_g$ (420 °C for 60 minutes) induced partial crystallisation of the amorphous structure, with the presence of both Cu-Zr intermetallic and Zr-oxide phases, which further increased $E$ and $H$ by acting as barriers to shear band propagation.

*Thermal Stability*

- The NLs exhibited higher thermal stability against crystallisation compared to their single components. The amorphous phase remained present up to 420 °C, whereas monolithic $Zr_{24}Cu_{76}$ and $Zr_{61}Cu_{39}$ films fully crystallised at 390 °C.
- The increased thermal stability is attributed to interdiffusion phenomena occurring at lower temperatures (from 300 °C), caused by the heterogeneous structure and interfaces of the NLs, which delay the onset of crystallisation. The evolution of the average composition toward $Zr_{50}Cu_{50}$, which has greater resistance against crystallisation due to the strong enthalpy of mixing, also plays a role.

Overall, we demonstrate that controlled thermal treatments can significantly influence the structural and mechanical properties of metallic glass nanolaminates, offering insights into their potential applications in advanced materials design.

**Acknowledgements**

Dr. M. Ghidelli acknowledges the financial support of the ANR MICRO-HEAs (Grant Agreement No. ANR-21-CE08-0003-01) as well as of the ANR "EGLASS" (Grant agreement no. ANR-22-CE92-0026-01). Moreover, he acknowledges the support from the Partenariats Hubert Curien (PHC) PROCOPE 2021 project ''New-Glasses" (Grant No. 46735ZG) financing the cooperation between LSPM-MPIE. Dr. J.P. Best acknowledges the German side of this scheme supported through the Deutsche Akademische Austauschdienst (DAAD) program "Programme des projektbezogenen Personenaustauschs (PPP)" (Project-ID: 57561649) financed by the Bundesministerium für Bildung und Forschung (BMBF). C. Jung is grateful for financial support from the Alexander von Humboldt Foundation. G. Dehm gratefully acknowledges support by the ERC Advanced Grant "GB Correlate" (Grant Agreement No. 787446). Benjamin Breitbach is thanked for assistance with setting-up the XRD measurements.

**Data Availability**

Data will be made available on request.

**Declaration of generative AI and AI-assisted technologies in the writing process**

During the preparation of this work the authors used ChatGPT (OpenAI) in order to improve the readability of some text sections. After using this tool/service, the authors reviewed and edited the content as needed and take full responsibility for the content of the published article.

# Thermal modification of ZrCu metallic glass nanolaminates: Structure and mechanical properties

Andrea Brognara[1], Chanwon Jung[1,a)], Cristiano Poltronieri[2,b)], Philippe Djemia[2], Gerhard Dehm[1,*], Matteo Ghidelli[2,*] and James P. Best[1,*]

[1] *Max Planck Institute for Sustainable Materials, Max-Planck-Str. 1, 40237 Düsseldorf, Germany*

[2] *Laboratoire des Sciences des Procédés et des Matériaux (LSPM), CNRS, Université Sorbonne Paris Nord, 93430 Villetaneuse, France*

[a)] *Present address: Department of Materials Science and Engineering, Pukyong National University, 45 Yongso-ro, Nam-gu, 48513 Busan, Republic of Korea*

[b)] *Present address: Carl Zeiss S.p.A., Via Varesina 162, 20156 Milan, Italy*

\* *Corresponding authors*: j.best@mpie.de (J.P. Best), matteo.ghidelli@lspm.cnrs.fr (M. Ghidelli), dehm@mpie.de (G. Dehm)

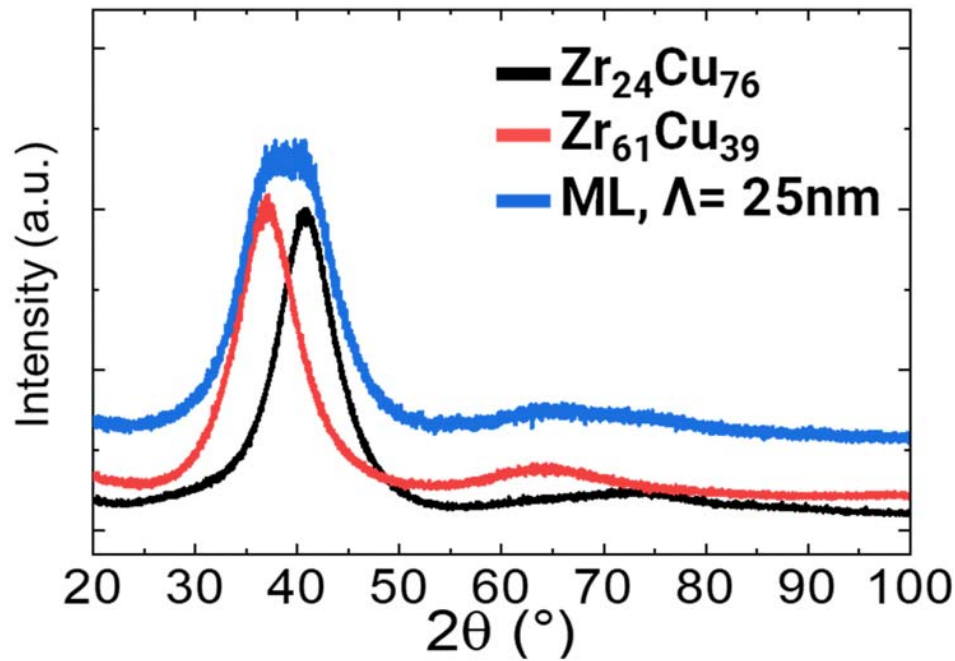

**Fig. S1**: Comparison of grazing incidence XRD diffractograms of $Zr_{24}Cu_{76}$ and $Zr_{61}Cu_{39}$ monolithic films and nanolaminate with $\Lambda$ = 25 nm (black, red, and blue curves, respectively).

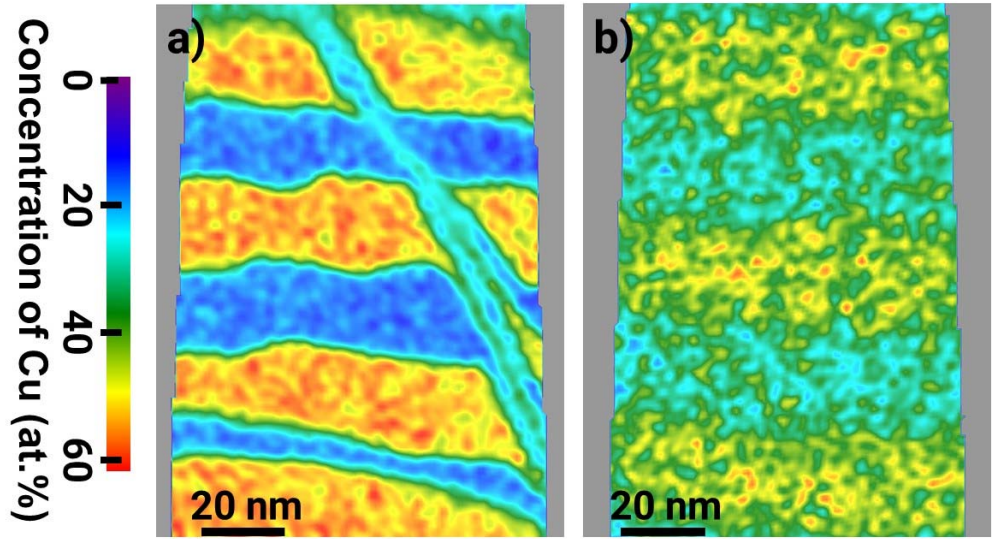

**Fig. S2**: Chemical concentration maps of the APT tips of the NL and thickness of 3 µm respectively in the (a) as deposited and (b) annealed at 330 °C for 1h conditions. The maps clearly show the variation of the layers chemical compositions after annealing, due to interdiffusion phenomena. The APT tip of the sample in the as deposited conditions (a) was extracted below the surface of a nanoindent and a SB was captured within its volume, as visible in the concentration map.

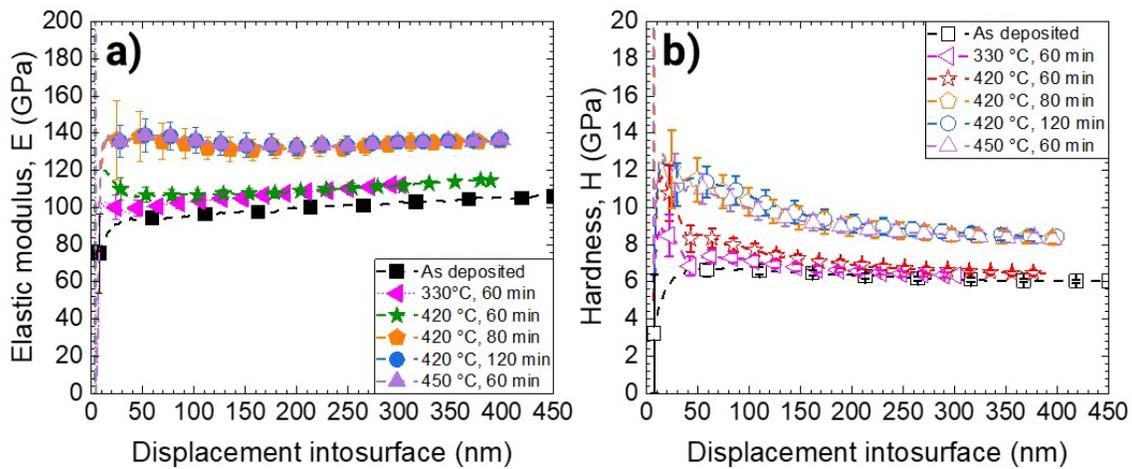

**Fig. S3**: CSM nanoindentation curves of a) elastic modulus ($E$) and b) hardness ($H$) as a function of indentation depth for the 3 µm thick NL in the as deposited conditions and after annealing at different temperatures and time. Both $E$ and $H$ gradually increase after annealing reaching a nearly constant value for the samples annealed at 420 °C for 80 and 120 min and at 450 °C for 60 min, indicating a full crystallisation.